\documentstyle[amssym,emulateapj]{article} 
\input{psfig}

\newcommand{\mincir}{\raise -2.truept\hbox{\rlap{\hbox{$\sim$}}\raise5.truept
\hbox{$<$}\ }}
\newcommand{\magcir}{\raise -2.truept\hbox{\rla669p{\hbox{$\sim$}}\raise5.truept
\hbox{$>$}\ }}
\newcommand{\minmag}{\raise-2.truept\hbox{\rlap{\hbox{$<$}}\raise 6.truept\hbox
{$>$}\ }}
\newcommand{\be}{\begin{equation}}
\newcommand{\ee}{\end{equation}}
\newcommand{\ba}{\begin{eqnarray}}
\newcommand{\ea}{\end{eqnarray}}
\newcommand{\brr}{\begin{array}}
\newcommand{\err}{\end{array}}
\newcommand{\bc}{\begin{center}}
\newcommand{\ec}{\end{center}}
\newcommand{\hm}{\,h^{-1}{\rm Mpc}}
\newcommand{\km}{\,h\, {\rm Mpc}^{-1}}
\newcommand{\etal}{{et al.}~}

\newcommand{\p}{\partial}
\newcommand{\f}{\frac}

\newcommand{\de}{\delta}

\newcommand{\eps}{\epsilon}
\newcommand{\s}{\sigma}
\newcommand{\al}{\alpha}
\newcommand{\lam}{\lambda}

\newcommand{\bfx}{{\bf x}}

\newcommand{\bfq}{{\bf q}}

\newcommand{\bfS}{{\bf S}}

\newcommand{\calL}{{\cal L}}
\newcommand{\calP}{{\cal P}}

\newcommand{\epsm}{\epsilon_{_{M}}}

\newcommand{\sigm}{\sigma_{_{\!M}}}

\newcommand{\0}{\circ}
\newcommand{\frakb}{{\frak b}}

\slugcomment{Ap.J.Letters, in press}

\lefthead{Catelan, Matarrese and Porciani}
\righthead{Halo spatial distribution}

\begin{document}

\title{ On the spatial distribution of dark matter halos}

\author{ Paolo Catelan\altaffilmark{1}, Sabino Matarrese
\altaffilmark{2} {\tiny AND} Cristiano Porciani \altaffilmark{3, 4}}
\affil{ \altaffilmark{1} Theoretical Astrophysics Center, Juliane
Maries Vej 30, 2100 Copenhagen \O, Denmark} 
\affil{ \altaffilmark{2} Dipartimento di Fisica `G. Galilei', 
Universit\`a di Padova, via Marzolo 8, 35131 Padova, Italy} 
\affil{ \altaffilmark{3} Scuola Internazionale Superiore di Studi 
Avanzati, via Beirut 4, 34014 Trieste, Italy} 
\affil{ \altaffilmark{4} Space Telescope Science Institute, 3700 San 
Martin Drive, Baltimore, MD 21218, USA}



\begin{abstract}
We study the spatial distribution of dark matter halos in the Universe
in terms of their number density contrast, related to the underlying
dark matter fluctuation via a non-local and non-linear bias random
field. The description of the matter dynamics is simplified by adopting
the `truncated' Zel'dovich approximation to obtain both analytical
results and simulated maps. The halo number density field in our maps
and its probability distribution reproduce with excellent accuracy
those of halos in a high-resolution $N$-body simulation with the same
initial conditions. Our non-linear and non-local bias prescription
matches the $N$-body halo distribution better than any Eulerian linear
and local bias.
\end{abstract}


\keywords{galaxies: statistics -- large-scale structure of Universe} 
 

%

\section{Introduction}

The simplest description for biasing assumes that the fluctuations in
the number density of luminous objects, $\de_{{\rm lum}}$, and in the
mass, $\de_{{\rm mass}}$, are proportional, $\de_{{\rm lum}} =
b\,\de_{{\rm mass}}$, where $b$ is the so called linear {\it bias
factor}.  Recently, Catelan \etal (1998, CLMP), following the seminal
papers by Cole \& Kaiser (1988) and Mo \& White (1996, MW), showed how
the relation between dark halos, recipient of the luminous matter, and
the underlying mass is to be cast in terms of a bias random {\it field}
$\frakb$, which depends in a non-local way on the mass density
field. Halo biasing is a process which evolves in time, depends on the
scales and the collapse times of the selected objects, but is
additionally determined by the gravitational conditions of the
environment. Most important, unlike previous models, CLMP treated halo
biasing as a non-local process.

In this Letter, we apply the CLMP bias model to analyze the spatial
halo distribution at several scales.  Mass particles move according to
the Zel'dovich (1970) approximation. We generally find excellent
agreement between our theoretical predictions and the distribution of
halos extracted from an $N$-body simulation with the same initial
conditions. In \S$\,2$ we present our bias model, in \S$\,3$ we test it
against simulations; \S$\,4$ contains our conclusions.

\section{The distribution of halo fluctuations}
\subsection{The model}

Let us consider a population of halos, selected in Lagrangian space
through their mass $M$ and formation redshift $z_f$. At any comoving
position $\bfx$ and observation redshift $z\leq z_f$, one can generate
Eulerian maps of the halo number density fluctuation $\de_h(\bfx,
z|M,z_f)$, given the mass density contrast $\de(\bfx, z)$ with
Lagrangian resolution $R_\0$ and the corresponding Lagrangian halo
density fluctuation field, $\de^L_h(\bfq|M,z_f)$, through the relation
\be \de_h(\bfx, z|M,z_f)= \left[ 1+\de^L_h(\bfq|M,z_f) \right] 
\left[1+\de(\bfx, z) \right] -1 \;, 
\label{deltah}
\ee
valid in the single-stream regime (CLMP).  The non-locality comes from
the fact that the halo number density in $\bfx$ is determined by its
initial value at the Lagrangian comoving position $\bfq$. Using a local
version of the Press \& Schechter (1974, PS) approach, we obtain
$$
\de^L_h(\bfq|M,z_f)=
\sqrt{2 \pi} \,\left[t_f-\eps_\0(\bfq)\right]\,\Theta
\left[t_f-\eps_\0(\bfq)\right]
$$
$$
\times\,\left\{\! \sqrt{\f{\pi}{2}}\f{t_f\Sigma}{\sigm}
\!\left[1+{\rm erf}\!
\left(\!\f{t_f\Sigma }{\sqrt{2}\s_\0\sigm}\!\right)\!\right]
+\s_\0\exp\!{
\left(\!-\f{t_f^2 \Sigma^2}{2 \s_\0^2 \sigm^2}\!\right)
}\!\right\}^{\!-1}
$$
\be
\times \,\f{\sigm^2}{\Sigma^2}\, 
\exp
\left[
-\f
{\eps_\0(\bfq)^2-2\eps_\0(\bfq)\,t_f+t_f^2\,\s_\0^2/\sigm^2}
{2\,\Sigma^2}
\right]
\;-\;1\,.\;
\label{colback}
\ee
This result is obtained as follows: adopting the peak-background split,
in the patch of fluid at $\bfq$ with Lagrangian size $R_\0$, one writes
the mean number-density of halos in terms of the PS formula, with a
local collapse threshold $t_f - \eps_\0$ modulated by the background
field $\eps_\0$, and finally removes the overall mean halo number
density [eq.(3) below].  Here $t_f\equiv\de_c /D(z_f)$, with $\de_c$
the critical threshold for collapse of a spherical perturbation and
$D(z)$ the linear growth factor of density fluctuations normalized to
unity at $z=0$ [in the Einstein-de Sitter universe, $\de_c \simeq
1.686$ and $D(z)=(1+z)^{-1}$]; $\eps_\0$ is the linear mass fluctuation
extrapolated to $z=0$ and smoothed on $R_\0$, with $\s_\0^2$ its
variance. Finally, $\sigm^2 = (2\pi^2)^{-1}\int_0^\infty
dk\,k^2\,P(k)\,W(kR)^2$, with $W(kR)$ the filter function and $P(k)$
the primordial power spectrum, is the variance on scale $M$ of the
linear density field $\epsm$ and $\Sigma^2 \equiv\sigm^2-\s_\0^2$.
Eq.(2) actually generalizes eq.(42) of CLMP in that collapse on the
background scale $R_\0$ is accounted for by the step function
$\Theta[t_f-\eps_\0]$: halos of mass $M \propto R^3$ cannot be present
in a collapsed region of Lagrangian size $R_\0>R$. As stressed by CLMP
(see also Porciani \etal 1998), this approach defines catalogs of halos
unaffected by the cloud-in-cloud problem (e.g. Bond \etal 1991) up to
the scale $R_\0$. By expanding eq.(\ref{colback}) to first order in
$\eps_\0$ one obtains, for $\sigma_\0 \ll \sigm$, $\de^L_h(\bfq|M,z_f)
\simeq b_1^L(M,z_f)\,\eps_\0(\bfq)$, where $b_1^L(M,z_f) \equiv
[t_f/\sigm^2-1/t_f]$ is the linear Lagrangian bias factor (MW).  Note
that in our approach the background scale $R_\0$ appears as a fitting
parameter which can be used to optimize the performance of the model.
As we will see in \S$\,3.2$, the optimal value of $\sigma_\0$ generally
depends on the chosen Eulerian resolution scale and on the masses of
the considered halo population.

When comparing with halos in numerical simulations, we will consider
finite mass intervals, so we will have to replace $\de_h^L$ in eq.(2)
with its weighted average, where the weight is given by the comoving
conditional mass-function
$$
n_h(M,z_f) = \f{\rho_b\;\exp{(-t_f^2/2\,\sigm^2)}} 
{2\pi\,M\,\sigm^2 \,\Sigma}\left|\f{d \sigm^2}{dM} \right|
$$
\be 
\times\left\{\!\sqrt{\f{\pi}{2}}\f{t_f\Sigma}{\sigm}
\!\left[1+{\rm erf}\!\left(\!\f{t_f\Sigma}
{\sqrt{2}\s_\0\sigm}\!\right)\!\right] +
\s_\0\exp\!{\left(\!-\f{t_f^2\Sigma^2}{2\s_\0^2\sigm^2}\!\right)}\!
\right\}, 
\ee 
with $\rho_b$ the mean density. In addition, as discussed by CLMP,
eq.(\ref{deltah}) can be generalized to multiple streaming as a
Chapman-Kolmogorov-type relation
\be
\de_h(\bfx, z|M,z_f)\!=\!\int\!\!d\bfq\,
\left[1+\de^L_h(\bfq|M,z_f)\right]\de_D[\bfx-\bfx(\bfq,z)]\,-1,
\label{chapkol}
\ee
with $\de_D$ the Dirac function: each fluid element of Lagrangian size
$R_\0$ carries a `halo density charge'
$n_h(M,z_f)\,[1+\de^L_h(\bfq|M,z_f)]$ along its trajectory.

At every point $\bfx = \bfq +\bfS$, with $\bfS$ the displacement of the
$\bfq$-th Lagrangian element smoothed on $R_\0$, we assign a halo
density $\de_h$ on the selected mass scale $M$ according to
eqs.(\ref{colback}) and (\ref{chapkol}). We extensively use this
formulation of our bias scheme in \S$\,3$ where we test the model
locally against a high-resolution $N$-body simulation.

\subsection{The distribution of halos} 

In this section we compute the probability distribution $p(\de_h)$
deriving from our bias model. We consider a mildly non-linear density
field in the laminar regime, though for comparisons with simulations we
will also adopt the multi-stream generalization in eq.(\ref{chapkol}).

Eq.(\ref{deltah}) can be recast, using mass conservation, in terms of
the Jacobian determinant $J\equiv |\!| \p \bfx/\p \bfq |\!|$ of the
mapping from Lagrangian to Eulerian space, $\bfq \rightarrow \bfx(\bfq,
z) =\bfq + \bfS(\bfq, z)\,$, namely $1+\de[\bfx, z]=J(\bfq, z)^{-1}$.
In the Zel'dovich approximation, $\bfS(\bfq,z) = - D(z) \nabla_\bfq
\varphi_\0(\bfq)$, where $\varphi_\0$ is the linear peculiar
gravitational potential, such that $\nabla^2_\bfq \varphi_\0(\bfq) =
\eps_\0(\bfq)$. The probability distribution of the eigenvalues
$\lam_\al(\bfq)$ ($\al=1,2,3$) of the deformation tensor
$\p^2\varphi_\0(\bfq)/\p q_\al\,\p q_\beta$ (Doroshkevich 1970) can be
used to compute the one-point statistical properties of $\de_h$ at any
redshift $z\leq z_f$ in Eulerian space.  Let us introduce the variables
$ \calL_\0\equiv \sqrt{5\,(\mu_1^2 - 3\mu_2)\,}\,, $ and $
\calP_\0\equiv(\lam_1-2\lam_2+\lam_3)/(2\sqrt{(\mu_1^2-3\mu_2)\,})\,, $
(Reisenegger \& Miralda-Escud\'e 1995) with $\mu_\al(\bfq)$
($\al=1,2,3$) the invariants of the deformation tensor. Unlike the
original eigenvalues, these variables are independent,
\be
p_{_{\!L}}\!(\eps_\0, \calL_\0,\calP_\0)\! = \!
\left(\f{{\rm e}^{-\eps_\0^2/2\s_\0^2}}{\sqrt{2\pi}\s_\0}\right)\!
\left(\f{2\,\calL_\0^4\,{\rm e}^{-\calL_\0^2/2\s_\0^2}\,}
{3\,\sqrt{2\pi}\s_\0^5}\right)\!
\left(\f{3}{2}-6 \calP_\0^2 \right), 
\label{RME}
\ee

 \centerline{{\vbox{\epsfxsize=8.3cm\epsfbox{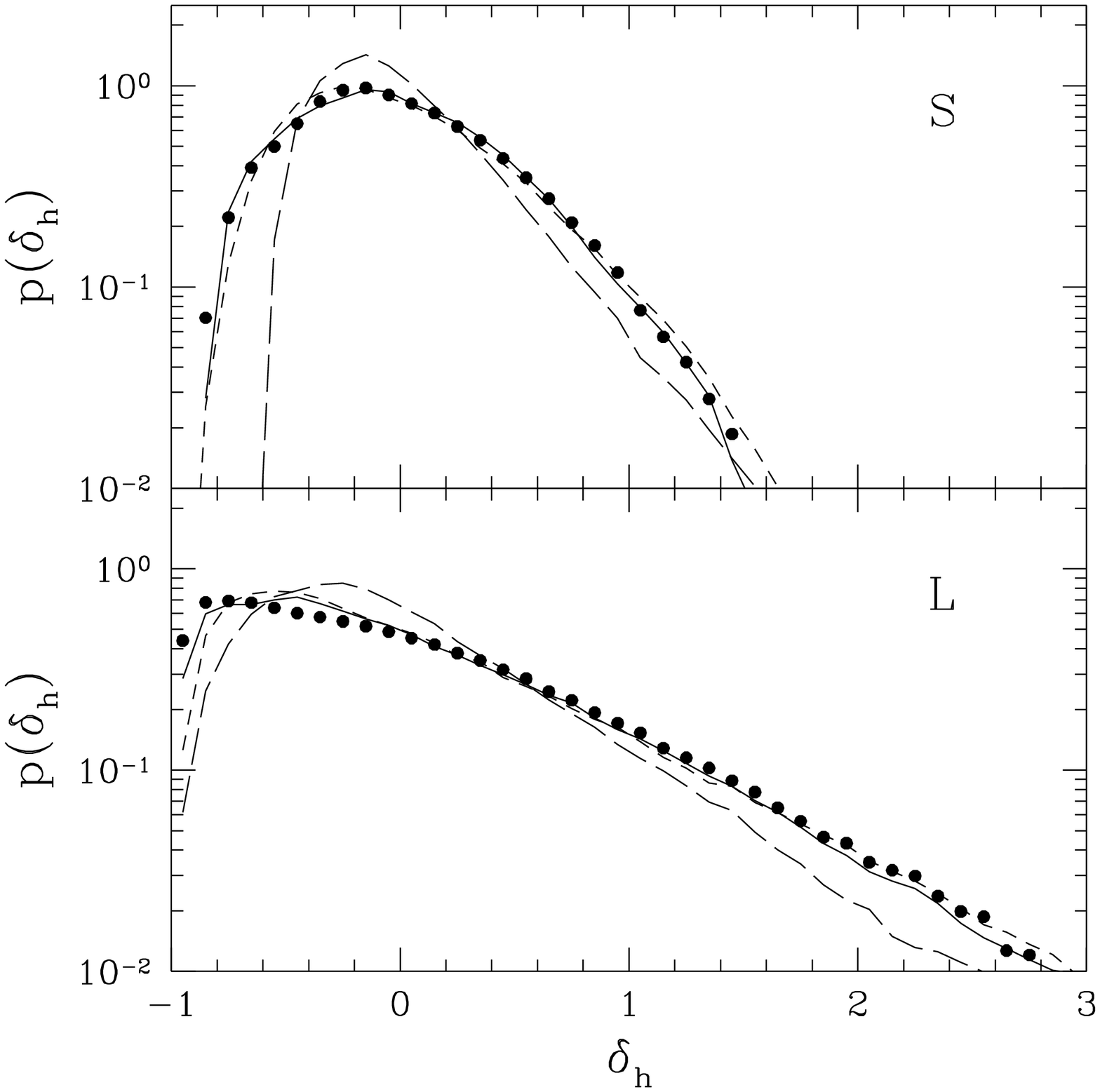}}}}
 {\footnotesize{Fig. 1.} --- The one-point distribution function of
 $\delta_h$. The short dashed lines refer to the data obtained by
 performing $10^5$ random generations of the variables $\eps_\0,
 \calL_\0$ and $\calP_\0$ ($k_\0=0.180 \km$, $\s_\0^2= 0.217$); the
 filled dots show the $N$-body output. The long dashed lines represent
 the linear bias prediction while the solid lines are the outcome of the
 numerical version of our model described in \S$\,3.2$. Top panels:
 class $S$ ($k_f= 0.314 \km$).  Bottom panels: class $L$ ($k_f=0.360
 \km$).}
 \vspace{0.3cm} 

\noindent where $0\leq\calL_\0\leq\infty$ and $-1/2 \leq \calP_\0\leq
1/2$.  The Jacobian now reads $J=1-D\{675\eps_\0 +
45D(\calL_\0^2-5\eps_\0^2) +
D^2[2\sqrt{5}\calL_\0^3\calP_\0(3-4\calP_\0^2)-15\calL_\0^2\eps_\0 +
25\eps_\0^3]\}/675$; it enters in eq.(1) through mass conservation.

The probability $p(\de_h)$ can then be computed by Monte Carlo
generating realizations from the distribution in eq.(\ref{RME}). Since
we are interested in the Eulerian probability and eq.(\ref{RME}) gives
a Lagrangian distribution, we compute $p(\de_h) = \int
d\de\;p_{_{\!L}}\!(\de_h,\de)/(1+\de)$, where $p_{_{\!L}}\!(\de_h,\de)$
is the joint Lagrangian probability for the Eulerian halo and mass
overdensity fields [cf. eq.(14) in Kofman \etal 1994]. In practice,
$p(\de_h)$ is obtained by: {\em i)} generating realizations for
$\eps_\0, \calL_\0, \calP_\0$; {\em ii)} computing $J$, and $\de_h$
through eq.(\ref{deltah}) or eq.(\ref{chapkol}); {\em iii)} weighting
the contribution to the probability of $\de_h$ by the factor $J$.

\section{Testing the model}

\subsection{Comparing the probability distribution function}

We test our predictions for $p(\de_h)$ against a high-resolution
$N$-body simulation from the data bank of cosmological simulations
provided by the Hydra Consortium and produced using the Hydra $N$-body
code (Couchman, Thomas \& Pierce 1995).  The simulation (RUN 501)
evolves $128^3$ particles on a $128^3$ cubic mesh with periodic
boundaries. The box size is $100 \hm$ and the particle mass $1.32
\times 10^{11} \,h^{-1}{\rm M}_\odot$.  The initial conditions are
Gaussian with a Cold Dark Matter spectrum with shape parameter
$\Gamma=0.25$, density parameter $\Omega=1$ and zero cosmological
constant. The simulation output corresponds to $\sigma_8=0.64$, where
$\sigma_8$ is the rms linear mass fluctuation in spheres of $8 \hm$.
At this epoch, the characteristic virializing halo mass, $M_\ast$,
defined by $\sigma_{_{M_\ast}}=\delta_c$, is $0.66 \times 10^{13}
\,h^{-1}{\rm M}_\odot$, i.e. 50.13 particles. To compare our
predictions to the $N$-body outcome we need a halo catalog from the
simulation. We adopt the spherical overdensity (SO) halo-finder (Lacey
\& Cole 1994) to identify spherical regions with mean overdensity
$\kappa=178$, leading to 5025 halos with more than 20 particles.  We
then consider two classes of objects: class $S$ contains halos with $
0.5 ~\mincir M/M_\ast \mincir 0.7 $; class $L$ has $3~\mincir M/M_\ast
\mincir 6$ (see Table 1). In Figure 1 we plot $p(\de_h)$ obtained with
our bias scheme against the $N$-body outcomes.  The model prediction
for $z=z_f=0$ is computed as in \S$\,2$.  $\sigma_\0$ has been tuned to
optimize the agreement with the numerical outputs.  The simulation
probability distribution has been extracted after smoothing the halo
distribution by a Gaussian filter $W = \exp(-k^2/2 k_f^2)$ with
resolution $k_f$. The prediction of a linear Eulerian bias model is
also plotted in Figure 1: the mass distribution in the simulation has
been smoothed with the same filter used for the halo overdensity and
the resulting $\delta$ is multiplied by the Eulerian bias factor
$b_{\rm MW}= 1 + [\de_c/\sigm^2 - 1/\de_c]$ (MW), reported in Table
1. Our model accurately reproduces the tail of the distribution for
positive $\de_h$, while for $\de_h<0$ it favours moderate
underdensities ($\de_h \sim -0.5$) with respect to the $N$-body
simulation.  The linear bias prescription instead produces much a less
skewed distribution with a higher peak and severely underestimates the
probability of very underdense regions.  Our model can be further
improved by adopting the multi-stream version introduced in \S$\,3.2$,
whose predictions, also plotted in Figure 1, are in excellent agreement
with the $N$-body outputs.

\subsection{Cross-correlations}

We also performed a much more severe point-by-point test, implementing
a fully numerical version of our bias scheme as follows. We consider a
computational box as large as the $N$-body one, but sampled with lower
resolution: $64^3$ particles on a $64^3$ grid (using $128^3$ particles
on a $128^3$ mesh gave identical results). Each particle is moved to
its final position according to the `truncated' Zel'dovich
approximation (Coles \etal 1993), that is, prior computing the
displacement, we remove initial power in high frequency modes by a
Gaussian filter with resolution $k_\0$; we require the amplitudes and
phases of the linear density field to be identical to the simulation
ones (at least for the Fourier modes present in both grids). Each
particle is then associated to the linear field $\eps_\0({\bf q})=
-\nabla_{\bf q} \cdot {\bf S}({\bf q})$.  In such a way, every particle
is endowed with its own halo-density charge $n_h \,[1+\delta^L_h({\bf
q})]$, computed as described in \S 2.1. Note that, while for
$\sigma_\0^2$ we used Gaussian smoothing, $\sigm$ is calculated with
top-hat filtering. The halo-density charge is carried by the particles
and eventually assigned to the $64^3$ grid through the Triangular
Shaped Cloud scheme. The corresponding halo overdensity field in the
multi-stream regime $\de_h^{\rm mod}$ is then computed and smoothed
with a Gaus-

\vspace{0.3cm}
%
%
 \centerline{
 \hbox{
 \psfig{figure=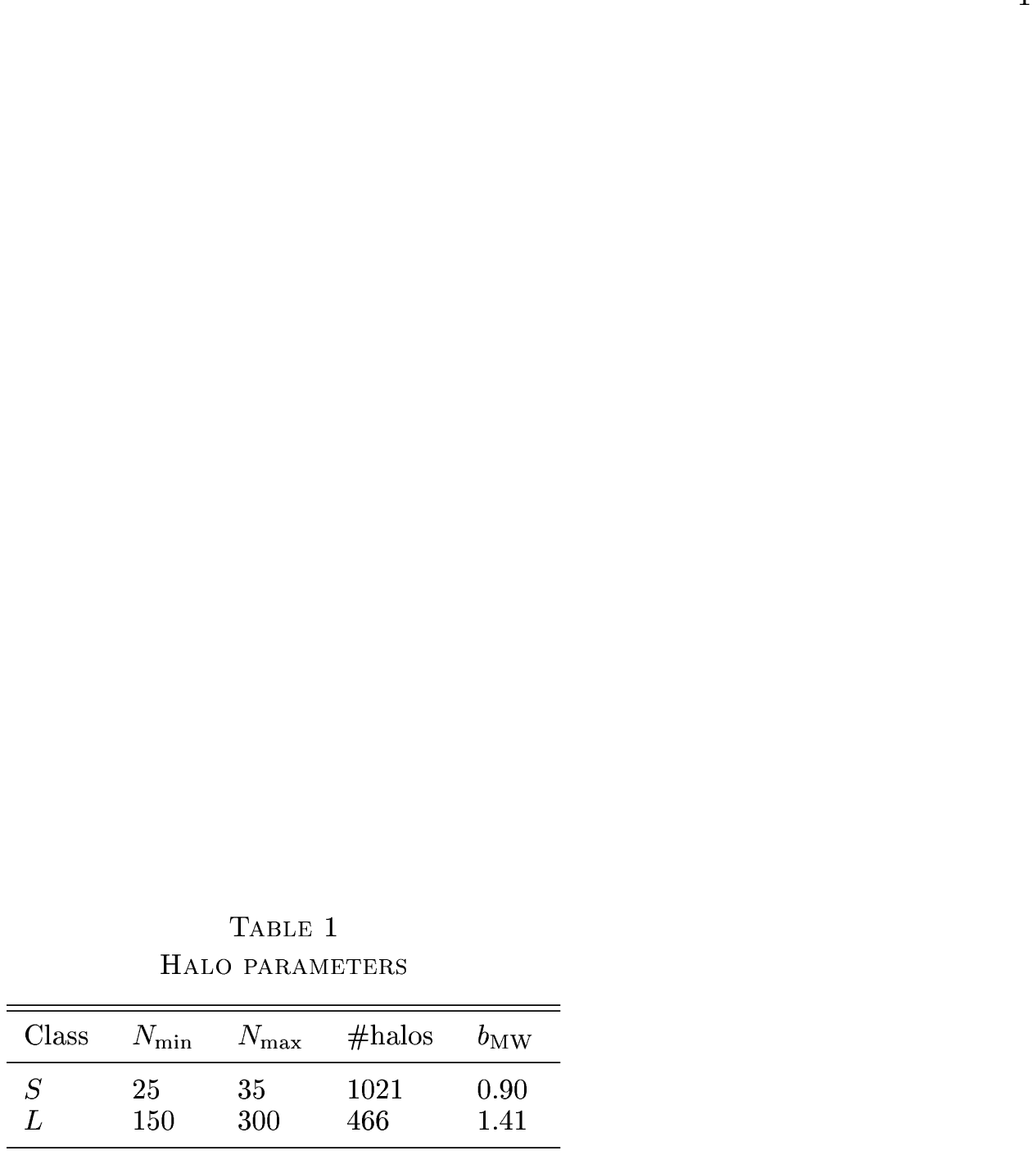,bbllx=212pt,bblly=388pt,bburx=406pt,bbury=473pt,height=2.8cm,width=7.0cm}
 }
 }

\centerline{{\vbox{\epsfxsize=8.3cm\epsfbox{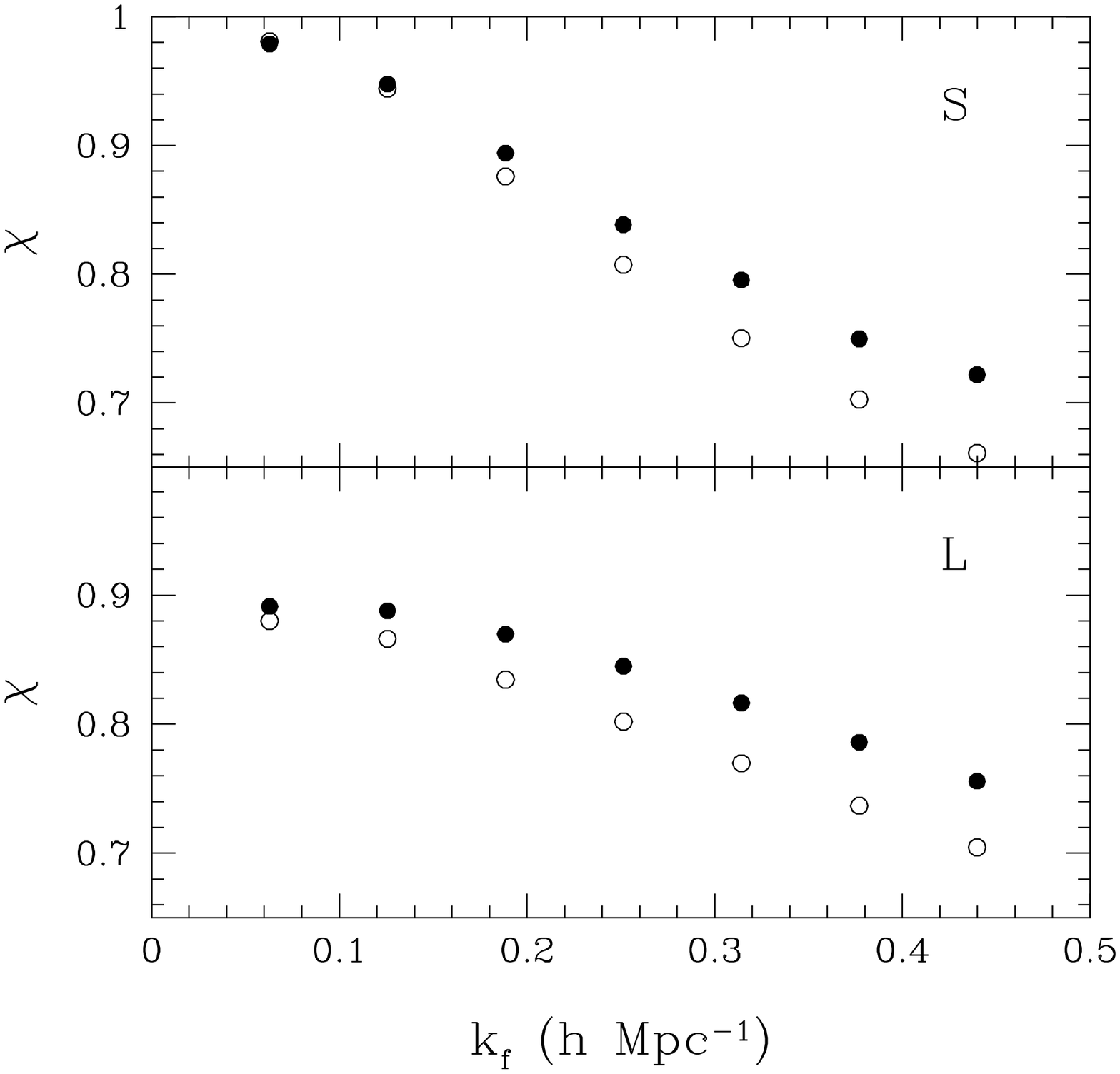}}}}
{\footnotesize{Fig. 2.} --- Cross correlation coefficients
vs. filtering scale. Filled and empty dots represent, respectively, our
model and any Eulerian linear bias scheme.}
\vspace{0.3cm}

\noindent sian filter with $k_f \leq k_\0$. The result has to be
compared with the halo field $\de_h^{\rm sim}$ extracted from the
N-body simulation, smoothed on the same scale. To quantify the
agreement between the two, we compute their cross-correlation
coefficient $\chi\equiv\langle\delta_h^{\rm mod}\, \de_h^{\rm
sim}\rangle/(\sigma_h^{\rm mod}\,\sigma_h^{\rm sim})$ where
$\sigma_h^i\equiv \langle\delta_h^{i\,2} \rangle ^{1/2}$, and the
average is performed over the grid points. A value $|\chi|=1$ means
that the two fields are proportional, while $\chi=0$ for uncorrelated
fields.  We tune the truncation $k_\0$ to optimize our bias scheme.
Since $\chi$ turns out to depend very weakly on $k_\0$ (in a wide range
around its optimal value), we can choose it so that $\sigma_h^{\rm mod}
= \sigma_h^{\rm sim}$ while keeping the maximum allowed value of
$\chi$. For each resolution this is obtained with $\s_\0^2 =
1.65\,\s_f^2+C $, with $C=1.45$ for class $S$ and $C=0.54$ for class
$L$. Note that in the MW model the value of $\sigma_h^{\rm
mod}/\sigma_h^{\rm sim}$ is not adjustable and generally differs from
$1$.  For the resolutions considered in Figure 3, $\sigma_h^{\rm
mod}/\sigma_h^{\rm sim}\,$ is $0.81$ for class $S$ and $0.90$ for class
$L$.  Moreover, considering smaller smoothing lenghts, this ratio can
differ from $1$ even by $30$\%.  This implies the MW method is unable
to accurately predict the average properties of the bias
distribution. Figure 2 reports $\chi$ values for the optimized model
vs. the resolution $k_f$.

To evaluate the performance of {\it any} Eulerian linear bias model, we
also show the cross-correlation coefficient when the smoothed mass
density field from the simulation is used instead of the output mass of
our model. This test does not depend on the value of the linear
bias. For each smoothing length our model reproduces the halo density
field of the simulation much more accurately than the linear biasing
scheme.  The performance of the two models is similar only for very
large smoothing lengths.

In Figure 3 we show the scatter obtained by plotting $\de_h^{\rm mod}$
vs. $\de_h^{\rm sim}$ for our model and for the linear bias scheme.  In
this case we adopt the MW bias factor in Table 1. Even though on
average our bias scheme gives better predictions (especially for class
$S$ and for underdense regions), some scatter in the relation
$\de_h^{\rm mod}$ vs. $\de_h^{\rm sim}$ persists. To test whe-

 \centerline{{\vbox{\epsfxsize=8.3cm\epsfbox{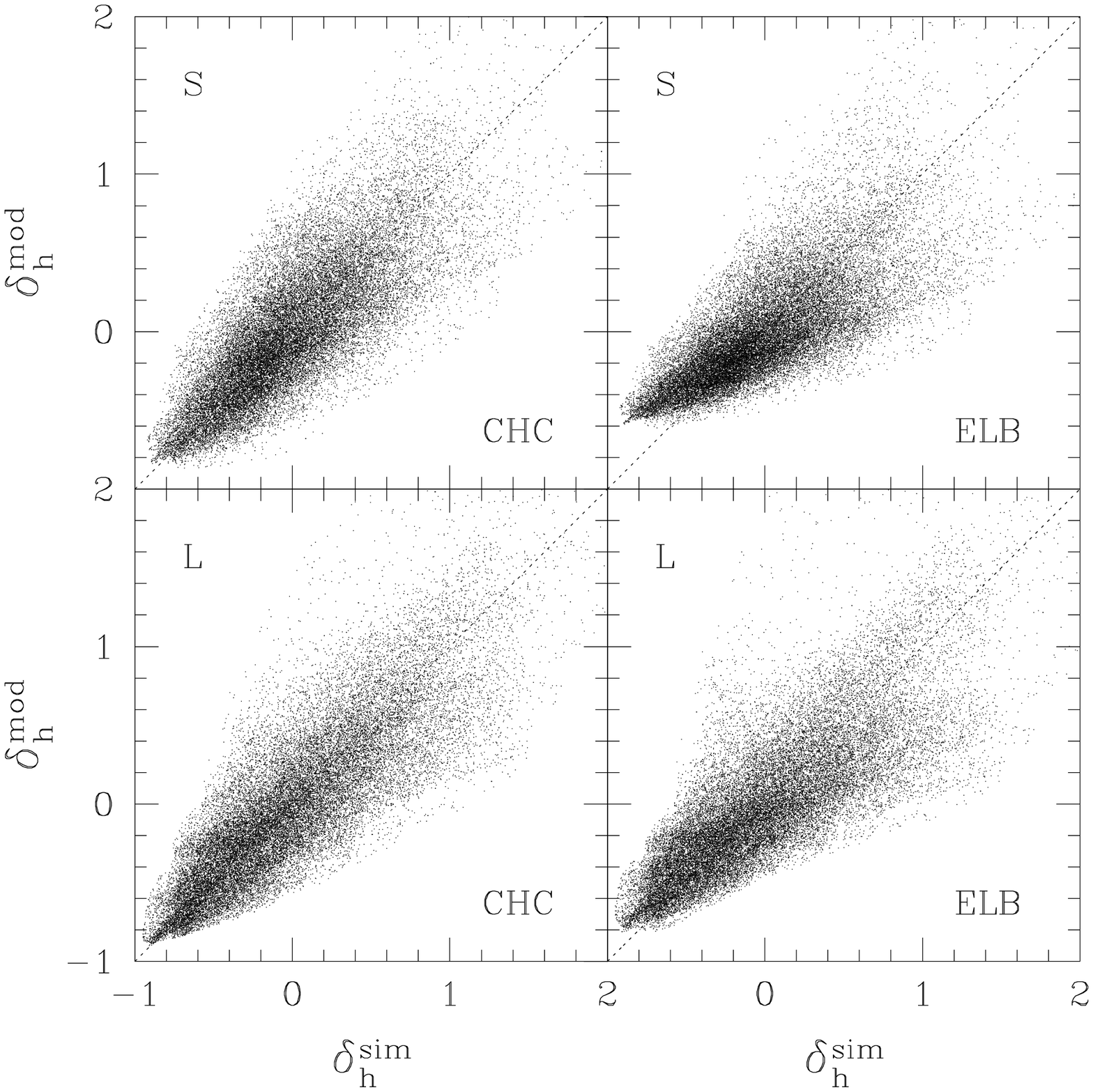}}}}
 {\footnotesize{Fig. 3.} --- Model predictions for $\de_h$ vs. $N$-body
 outcome.  Left panels: our bias scheme (Conserved Halo-Density Charge,
 CHC). Right panels: Eulerian Linear Bias model (ELB) in the MW version.
 Top panels: class $S$ (with $k_f=0.260 \km $).  Bottom panels: class
 $L$ (with $k_f=0.301 \km$). A 1-in-8 random selection is shown.}
 \vspace{0.3 cm}

\noindent ther this is due to our simplified dynamics, we generate new
halo maps taking the particle displacements directly from the $N$-body
simulation. The results are in excellent agreement with those obtained
with the Zel'dovich approximation, indicating that the local PS
approach, being unable to accurately model the Lagrangian halo counting
(Fig. 8 in White 1996), is actually responsible for the scatter. We
will address this point in a future work.

\section{Discussion and conclusions}

We devised a simple and fast semi-analytical technique which allows to
study the spatial distribution of dark matter halos in terms of their
local number density contrast. Our method, which is based on a
Lagrangian halo identification algorithm plus the Zel'dovich
approximation for the matter dynamics, was successfully tested against
the distribution of halos extracted from a high resolution N-body
simulation. Possible improvements should go in the direction of
refining the halo selection criterion in Lagrangian space, e.g. using
the ellipsoidal collapse model or the peak theory as in (Bond \& Myers
1996; see also Monaco 1998).

As stressed by CLMP, our model can be applied to study the evolution of
galaxy biasing, once the relation between the galaxies and the hosting
dark matter halos is specified (e.g. Matarrese \etal 1997).  In
particular, defining the bias field such that $\de_h\equiv
\frakb\,\de$, from eq.(7) one obtains $\frakb[\bfx(\bfq,z),
z]=1+\de_h^L(\bfq|z_f)/[1-J(\bfq,z)]$. Tegmark \& Peebles (1998) have
recently stressed the importance of the asymptotic trend of the bias
factor. We can analyze this issue in the present context by considering
a galaxy population conserved in number after an initial merging phase
(i.e. for varying $z$ at fixed $z_f$). In the Einstein-de Sitter case
we recover the `debiasing' predicted by linear theory: $\frakb \to 1$
as $z\to -1$. Differently, if $\Omega<1$, $\frakb$ tends to a
space-dependent value which generally differs from 1; linear theory
would predict $b\to 1+(b_0 - 1)/D_{-1}(\Omega_0)$, as $z\to -1$, with
$b_0\equiv b(z=0)$ and $D_{-1}\equiv D(z=-1)$.

Our method can be used to analyze the coarse-grained statistical
properties of galaxies and clusters at various redshifts, e.g. applying
semi-analytical techniques to relate the dark matter halo distribution
to that of luminous objects like galaxies (e.g. Kauffmann, Nusser \&
Steinmetz 1997). It can be further implemented to generate very large
mock maps of these objects in our past light-cone, a problem made of
compelling relevance by the ongoing wide-field redshift galaxy surveys
like the 2 Degree Field Survey (2dF) and the Sloan Digital Sky Survey
(SDSS). Maps of the $X$-ray cluster distribution may also be produced
by the present method.

\vspace{0.3cm} We thank S. Cole and C. Lacey for providing the SO halo
finder and the Hydra Consortium for $N$-body simulations
(http://coho.astro.uwo.ca/pub/data.html).  PC has been supported by the
Danish NRF at TAC; SM and CP by the Italian MURST. CP thanks support
from NASA ATP-NAG5-4236 grant.

\end{document}